\documentstyle[twoside,fleqn,espcrc2,epsfig]{article}

\newcommand{\AmS}{{\protect\the\textfont2
  A\kern-.1667em\lower.5ex\hbox{M}\kern-.125emS}}

\def\ltsima{$\; \buildrel < \over \sim \;$}
\def\simlt{\lower.5ex\hbox{\ltsima}}
\def\gtsima{$\; \buildrel > \over \sim \;$}
\def\simgt{\lower.5ex\hbox{\gtsima}}

\title{Beppo-SAX Observations of Galaxy Clusters}

\author {S.Molendi\address{IFCTR, CNR, Milano,Italy}}

\begin{document}

\begin{abstract}
The high spatial resolution of the MECS experiment on board Beppo-SAX 
has encouraged a few scientists, including the author, to perform observations
of galaxy clusters. Results from the analysis of the first few observed 
objects are encouraging. After having reviewed the 
Beppo-SAX observing program for galaxy clusters and referenced contributions 
to these proceedings by other authors on the same topic, I present results
from the analysis of the Perseus cluster.

\end{abstract}

\maketitle

\section{Introduction}
Although principally devoted to studies of compact galactic sources and active galactic nuclei,
the Beppo-SAX satellite has devoted a small fraction of time to galaxy
cluster observations. The primary motivation for performing observations of 
galaxy clusters with Beppo-SAX is the high spatial resolution of the MECS 
instruments. 
In these proceedings I briefly review the Beppo-SAX observational program for 
galaxy clusters. After having listed contributions on galaxy clusters to these
proceedings by other authors I describe results from the analysis of the Perseus 
observation.  
  
\section{Instrumental Issues}

The MECS instrument on board Beppo-SAX is well suited for 
the analysis of extended sources. As discussed in Giommi et al. (1998) in 
these proceedings, the effective area of the 3 MECS units is comparable 
to that of the 2 GIS units on board ASCA in the 2-10 keV (see their Figure 1).
The Point Spread Function of the MECS is significantly sharper than that 
of the ASCA GIS (see Giommi et al. 1998 Figure 2). At 6.4 keV the half
power radius (HPR) of the MECS PSF is  $\sim$ 1 arcmin while the 
HPR of the ASCA GIS PSF is  $\sim$ 2 arcmin. Moreover, unlike the 
ASCA GIS PSF, the MECS PSF does not vary strongly with energy (see Fig.1
of D'Acri et al. 1998, these proceedings). 
Another important point is that the  instrumental background for the 
MECS units on-board Beppo-SAX is quite low   
$\sim 2.3\times 10^{-5} cts/s/arcmin^2/unit $ in the 1.5-10 keV band
to be compared with
$\sim 4\times 10^{-5} cts/s/arcmin^2/unit $ in the same band for the GIS 
units on board ASCA.
Moreover, due to the low inclination orbit adopted by the Beppo-SAX satellite,
the background is extremely stable.

\section{Beppo-SAX clusters observations}
The Beppo-SAX observing program includes 
observations of a few galaxy cluster.  A list of observations scheduled 
and performed 
is presented in these proceedings by Pallavicini (see his Table 3).
Almost all these observations are devoted to extended objects, the main aim 
being spatially resolved spectroscopy.
Within these proceedings Beppo-SAX observations of galaxy clusters 
are presented by a few authors. Kaastra et al. (1998) discuss spatially
resolved spectroscopy of A2199. They find evidence of a spectral hardening in 
the outer parts of this cluster. Moreover, using PDS data they confirm the presence 
of a hard tail in A2199. Colafrancesco et al. (1998)
discuss the analysis of 2 intermediate redshift clusters namely 
A33 and A348. They combine their data with previous ASCA measurements
to place constraints on the $L_x - T$ relationship at $z \sim 0.3$.
D'Acri et al. discuss an analysis technique which they have developed to 
correct for the spectral distortions introduced by the energy dependent 
MECS PSF.
They apply this technique to the core of the Virgo  cluster finding 
a temperature decrement towards the inner parts of cluster, in agreement with 
previous measurements made with ROSAT and ASCA.



\section{Perseus}

The central region of the Perseus cluster was observed by the Beppo-SAX 
satellite (Boella, Butler et al. 97) 
between the 20$^{th}$ and the 21$^{st}$ of September 1996
during the Science Verification Phase (SVP).
The observation was rather long with a total effective exposure of 89 ks 
for the Medium Energy Concentrator Spectrometer (MECS) 
(Boella, Chiappetti et al. 97), 38 ks for the 
HPGSPC instrument (Manzo et al. 1997) and 30 ks for the PDS
instrument (Frontera et al. 1997). 

\subsection{Hard Component}

\begin{figure}[t]
\vspace {-0.3truecm}
\epsfig{figure=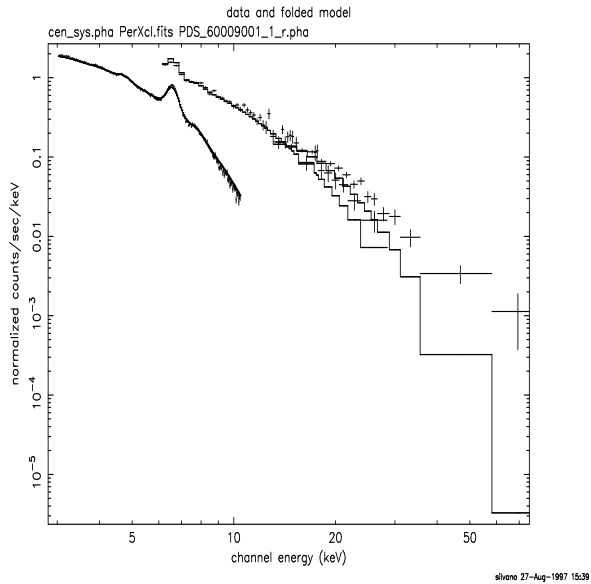,height=8cm, width=8cm, angle=-90}
\vspace {-1.0truecm}
\caption{The MECS, HPGSPC and PDS Spectrum of the central region of Perseus. 
The data is fitted with a thermal emission model. The strong excess 
above 25 keV is most likely due to the AGN in NGC 1275}
\vspace {-0.3truecm}
\end{figure}

I have accumulated a  MECS spectrum from a region centered
on the emission peak using an extraction radius of 6.4 arcmin.
I have fitted the MECS, HPGSPC and PDS spectra simultaneously using a 
thermal emission code  (MEKAL in XSPEC version 9.01) absorbed by 
an equivalent hydrogen column density of  N$_{\rm H}= 1.39\times
10^{21}$ cm$^{-2}$ 
The fit to the broad band spectrum (Figure 1) shows an excess
of the data with respect to the model at energies E \gtsima 20 keV.
This excess cannot be fitted with a thermal component, unless an
unphysical temperature of more than 20 keV is assumed. A power-law
component fits adequately the high energy excess. The intensity of 
the power-law component is $\sim 4\times 10^{-11}$ erg cm$^{-2}$s$^{-1}$ in
the 20-100 keV band. Comparison of our measurement with previous
detections by OSO7 (Rothschild et al. 1981), HEAO1 (Levine et al. 1984)
and with the OSSE upper limit (Osako et al. 1994) indicate that the 
power-law component is undergoing a secular decrease in intensity. 

\subsection{Cooling Flow}

\begin{figure}[t]
\vspace {-0.3truecm}
\epsfig{figure=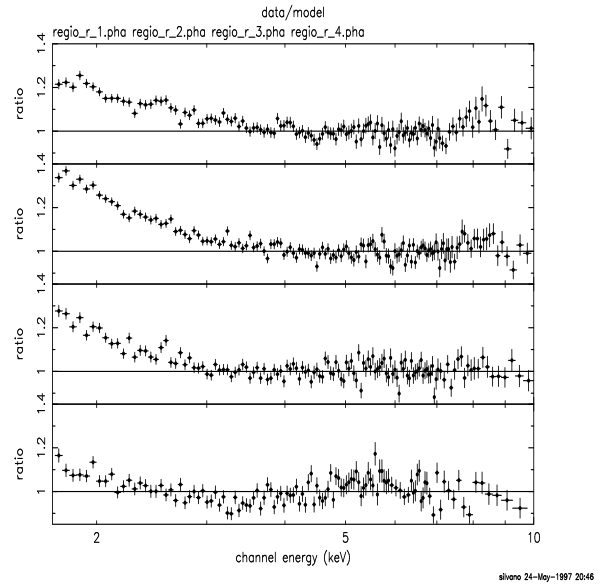,height=8cm, width=8cm, angle=-90}
\vspace {-1.0truecm}
\caption{Residuals in the form of ratio of data to model for 4 MECS spectra 
accumulated in concentric annuli centered  on the emission peak and with 
bounding radii of 0-2 arcmin, 2-4 arcmin, 4-6 arcmin and 6-8 arcmin 
respectively. Each spectrum is fitted with a thermal model in the 
energy range 3.5-10 keV, data below 3.5 keV is re-included after the 
fitting process.}
\vspace {-0.3truecm}
\end{figure}

An ASCA analysis of the central region of Perseus (Fabian et al. 94)
shows that the spectrum below 3 keV is strongly contaminated by a cooling
flow. By comparing the surface brightness profile of Perseus in the
0.5-2.0 keV band, derived from the analysis of ROSAT PSPC archive data, with 
a King profile we find that the cooling flow extends out to $\sim$ 8 arcmin, 
corresponding to 240 kpc. 
I have attempted to spatially resolve the cooling flow component 
with the MECS by analyzing spectra from 4 concentric annuli centered
on the emission peak and with bounding radii of 0-2 arcmin, 2-4 arcmin,
4-6 arcmin and 6-8 arcmin respectively.  I have fitted each spectrum
in the 3.5-10 keV range with a thermal model (MEKAL) allowing the 
temperature and the abundances to be free. I have then reintroduced 
the data in the 1.5-3.5 keV range and plotted the residuals to the fits
(see Figure 2).
As can be clearly seen the relative intensity of the cooling flow 
component decreases as the radial distance from the peak of the
emission increases (the increase observed when going from panel 1
to panel 2 is due to instrumental effects, more specifically 
the detector PSF, below 3 keV, redistributes a large fraction 
of photons coming from a circular region with a radius of 2 arcmin 
over a larger region). 

\begin{figure}[t]
\vspace {-0.5truecm}
\epsfig{figure=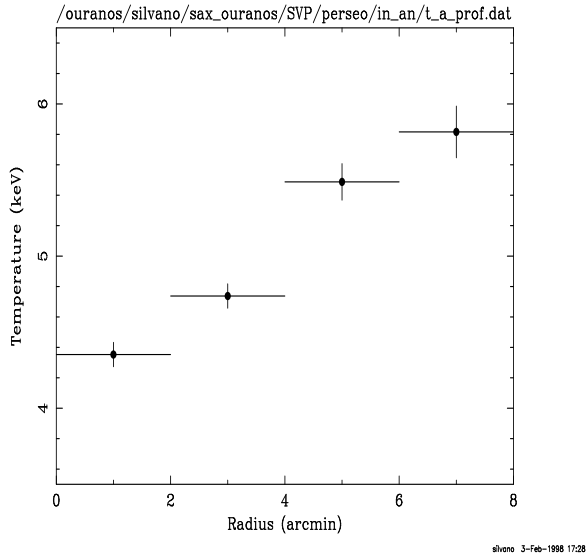,height=8cm, width=8cm, angle=-90}
\vspace {-1.0truecm}
\caption{Temperature versus radius. Reported values  for of 0-2 arcmin, 
2-4 arcmin , 4-6 and 6-8 arcmin have been obtained by performing spectral 
fitting of spectra in the energy range 3-8 keV.}
\vspace {-0.3truecm}
\end{figure}

\subsection{Temperature Gradient}

As described in the previous subsection I have fitted spectra 
accumulated in concentric annuli with a thermal emission model.
In Figure 3 I plot the temperature derived from 
the fits. In order to avoid contamination from the cooling flow the 
fits were performed in the 3-10 keV band.
Our data clearly indicates that the temperature presents a  strong gradient
in Perseus over a region of 8 arcmin (corresponding to 240 kpc) in radius. 

\subsection{Line Ratios}

I have accumulated a  MECS spectrum from a circular region centered
on the emission peak using an extraction radius of 6.4 arcmin, corresponding
to $\sim$ 200 kpc.
I have modeled the spectrum in the 3-10 keV band using a bremsstrahlung for
the continuum and two Gaussian lines, one for the 6.8 keV iron complex,
the other for the 8 keV iron-nickel complex.
Data at energies 
below 3 keV has been excluded because significantly contaminated by the 
cooling flow.  
I find a temperature of 
$4.9\pm 0.1$ keV for the continuum component, an energy E$_1=6.76\pm 0.01$ 
keV and a width $\sigma_1 = 0.04\pm 0.04$ keV for the lower energy line 
and an energy E$_2=8.06\pm 0.04$ keV and a width $\sigma_2 =0.30\pm0.05$ 
keV for the higher energy line, where both E$_1$ and E$_2$ are given in
the source rest-frame.   
The 8 keV feature is clearly broad and most likely 
due to a blend of different lines from highly ionized 
iron (Fe XXV and Fe XXVI) and nickel (Ni XXVII). 

\begin{figure}[tb]
\vspace {-0.9truecm}
\epsfig{figure=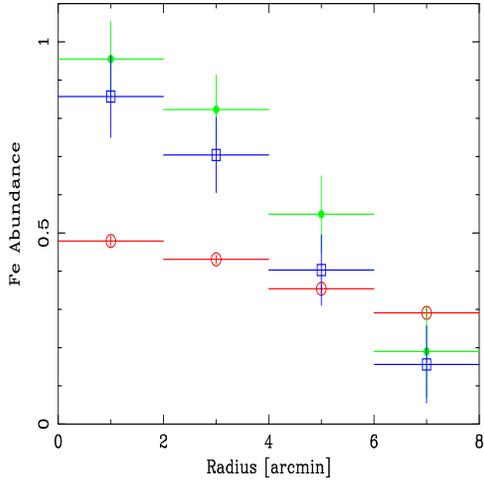,height=8cm, width=8cm, angle=-90}
\vspace {-1.0truecm}
\caption{Iron radial abundance profile. The ``apparent'' abundances,
measured using the Fe K$_{\alpha}$ line under the assumption of 
optically thin emission, are indicated with the open circles.
The filled circles indicate the abundances obtained by correcting 
the ``apparent'' abundances. The open squares
indicate the abundances obtained using the Fe K$_{\beta}$ line.}
\vspace {-0.3truecm}
\end{figure}

Fits with thermal emission codes such as MEKA, MEKAL
or Raymond \& Smith do not reproduce satisfactorily the 8 keV feature,
all these models underestimate significantly the intensity of the 
emission feature, 
the reason being that all the above codes predict an 8 keV over 6.8 keV
line intensity ratio in the range 0.11-0.13 against an observed ratio of 
$0.20\pm 0.02$. 
The most convincing explanation I have found 
for the observed anomalous ratio involves a process known as  
resonant scattering (a detailed discussion of alternative explanations
and of other aspects of the line ratio analysis is presented in Molendi 
et al. 98). 
The inter-galactic medium in clusters,
while optically thin to the continuum and many lines, may be optically
thick in the center of some lines. 
Indeed the process describing the absorption of an 
Fe K$_{\alpha}$ line photon by an iron ion followed by the 
immediate re-emission, which is known as resonant scattering, can be quite 
effective for typical cluster gas 
densities and temperatures. Gilfanov et al. (86) have shown that
the cores of rich cluster, such as Perseus and Virgo, should have 
optical depths, for the above process, of the order
of a few. 

If the gas is optically thick to resonant scattering, 
the line emission coming from the core of the cluster will be attenuated
because of the photons which are scattered out of the line of sight. 
If this is the case then the abundances
measured using standard thermal emission codes, which
assume optically thin thermal emission, can be significantly 
underestimated. In Figure 4 I show the abundance profiles computed 
under the assumption of optically thin line emission (open circles)
and of optically thick emission, using 2 different methods (filled circles
and open squares). I find that the the Fe abundance in the innermost 
circular region, with radius 2 arcmin,  
corresponding to $\sim$ 60 kpc,  is  $\sim$ 0.9 solar and consistent with 1.
The obvious implication is that a very large fraction of the gas in 
the core of Perseus has been processed in stars.  

\section{Acknowledgments}
I acknowledge support of various nature from my colleagues in the 
Beppo-SAX team.
I am particularly grateful to the LOC for the exquisite choice of restaurant for 
the social dinner.


\begin{thebibliography}{9}

\bibitem{Boe97a} Boella, G., Butler, C. et al.  1997 A\&AS, 122, 299
\bibitem{Boe97b} Boella G., Chiappetti, L. et al. 1997 A\&AS, 122, 327
\bibitem{} Colafrancesco, S., Antonelli, A., et al. {\it these proc.} 
\bibitem{} D'Acri F., De Grandi, S., Molendi, S., {\it these proc.} 
\bibitem{} Fabian A. C.; Arnaud, K. A.; et al. 1994, ApJL 436 63
\bibitem{} Frontera, F., Costa, E. et al. 1997, A\&AS 122 357 
\bibitem{} Gilfanov, M. R., Syunyaev, R.,A., et al. 1986, 
Soviet Astronomy Letters, 13, 1
\bibitem{} Giommi, P., Fiore, F., et al. {\it these proc.} 
\bibitem{} Kaastra, J., Bleeker, J.A.M, Mewe, R. {\it these proc.} 
\bibitem{} Levine, A. M.; Lang, F. L.; et al.
1984, ApJS 54 581
\bibitem{} Manzo, G., et al. 1997, A\&AS 122 341
\bibitem{} Molendi, S. , Matt, G., et al. 1998 ApJ 
in press 
\bibitem{} Osako, C.Y.; Ulmer, M. P.; et al. 1994 
ApJ 435 181
\bibitem{} Pallavicini {\it these proc.} 
\bibitem{} Rothschild, R. E.; Baity, W. A.; et al. 1981, ApJ, 243 9
\end{thebibliography}
\end{document}